\documentstyle[prb,aps,epsf,epsfig]{revtex}
\newcommand{\be}{\begin{equation}}
\newcommand{\bff}{\rm}
\newcommand{\ee}{\end{equation}}
\newcommand{\bea}{\begin{eqnarray}}
\newcommand{\eea}{\end{eqnarray}}

\begin{document}  
\twocolumn[\hsize\textwidth\columnwidth\hsize\csname@twocolumnfalse\endcsname 
\title{Inversion of ARPES measurements in high T$_c$ cuprates}
\author{S. Verga, A. Knigavko, and F. Marsiglio}
\address{Department of Physics, University of Alberta, Edmonton,
Alberta, Canada T6G 2J1}
\maketitle 

\begin{abstract} 
Recent energy dispersion measurements in several families of the hole-doped
copper oxides have revealed a kink in the energy vs. momentum relation. These
have tentatively been identified as due to electron phonon coupling. We
invert this data directly to determine the bosonic spectral function; the kink
gives rise to a singular function in the phonon energy region.
\end{abstract}


\vskip1pc]
 
The determination of mechanism for superconductivity in the high
temperature oxides has occupied researchers for the past fifteen years.
The most definitive signature for determining mechanism in conventional
superconductors traditionally has been the measurement of the single
particle tunneling I-V characteristic \cite{rowell63}, and the concomitant
inversion procedure \cite{mcmillan65}. Researchers 
\cite{tsuda00} have reported some success with this procedure
for the high temperature superconductors; nonetheless the applicability of
such a procedure is unknown outside a weak coupling electron phonon
framework, and the process is complicated in the superconducting state due
to the non-isotropic nature of the order parameter. Other procedures
have been suggested, such as the inversion of the normal state optical
conductivity \cite{marsiglio98}, the conductivity in the superconducting
state in conjunction with neutron scattering data \cite{carbotte99},
and the inversion of photoemission data \cite{arnold91} in the superconducting
state. Recent very high resolution photoemission measurements on a variety
of cuprate materials have suggested that a sizeable electron phonon coupling
exists \cite{lanzara01}, and the possibility of inverting this data (in the
normal state) has been re-opened. 
\begin{figure}[tb]
\begin{center}
\epsfig{figure=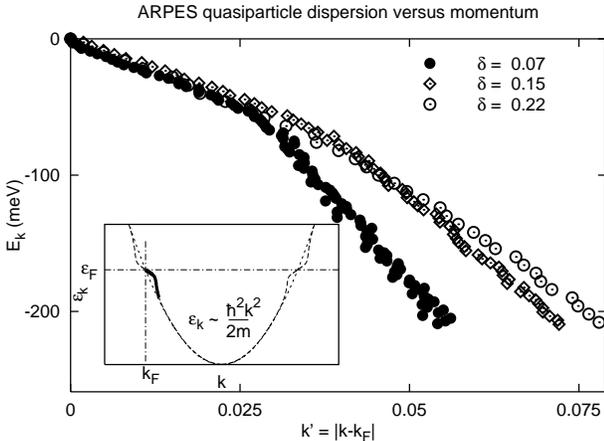,width=8cm}
\caption{
Dispersion of occupied states vs. momentum
for LSCO with three different doping concentrations. The insert shows a
schematic which puts this data in the context of a parabolic band. A modification
due to electron phonon coupling occurs near the Fermi surface. The region
corresponding to the data of the main figure is highlighted in bold on the
left side of the parabola.
}
\end{center}
\end{figure}
In this paper we outline an inversion
procedure, determine the `bosonic spectrum' to which the 
electrons are coupled, and assess the possibility that these bosons are
the phonons. The data for the dispersion for three different dopant concentrations
in LSCO is reproduced in Fig. 1 \cite{lanzara01}.
Particularly for the underdoped
sample there is a well-defined kink which occurs at approximately 70 meV. Lanzara
{\it et al.} \cite{lanzara01} attributed this kink to an electron self energy effect due to coupling
to phonons. We wish to investigate this claim based on some microscopic models.

We first examine the result obtained from `standard' phonon models, namely the
Einstein and Debye models. Each in turn is used to model the electron phonon
spectral function, $\alpha^2F(\nu)$, and then, within the standard framework
\cite{englesberg63,allen82}, the electron self energy $\Sigma(\omega) \equiv 
\Sigma_1(\omega) + i \Sigma_2(\omega)$ is obtained (at $T = 0$):
\begin{equation}
\Sigma_1(\omega + i\delta) = \int_0^\infty d\nu  \
\alpha^2F(\nu) 
\log{\left|{\omega - \nu \over \omega + \nu}\right|}
\label{realself}
\end{equation} 
\noindent from which the electron dispersion $E_k$ can be obtained:
\begin{equation}
E_k = \epsilon_k + \Sigma_1(E_k),
\label{self}
\end{equation}
\noindent where $\epsilon_k$ is the bare (with respect to electron phonon interactions)
quasiparticle energy. The model spectrum, along with the real part of the self energy and the
dispersion is plotted in Fig. 2(3) for the Einstein (Debye) models, respectively.
{\bff Note that for some values of $\epsilon_k$ the dispersion as shown is multivalued.
This is because multiple poles exist in these regions. In fact, the spectral function
generally evolves in the following manner \cite{englesberg63} (we use the Einstein model
depicted in Fig. 2 for simplicity): at low energies a pole exists just above the real
axis (i.e. with infinitesmal width). What is not depicted in Fig. 2 or 3 is that the weight
of this pole (i.e. the residue) goes to zero as $|\epsilon_k| \rightarrow \infty$. So, 
beyond about $|\epsilon_k| \approx 140$ meV, the weight of this pole becomes very small. The
actual energy ($E_k$) at this point becomes very nearly the Einstein frequency. In a sense the
electrons and phonons have `hybridized' and this branch, which started out very `electron-like'
near $E_k \approx 0$ is now very 'phonon-like'. The branch just below -80 meV in Fig. 2c is
more or less irrelevant, since the self energy has a very large imaginary part (see Fig. 2b).
Finally, as $|\epsilon_k|$ continues to increase the lowest branch begins to dominate. In this 
limit this branch becomes very `electron-like' albeit with a finite width.}

\begin{figure}[tb]
\begin{center}
\vskip0.2cm
\epsfig{figure=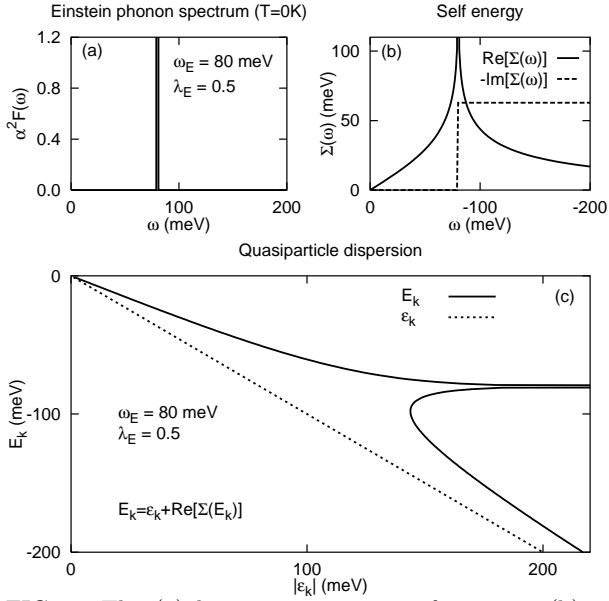,width=8cm}
\caption{The (a) boson spectrum vs. frequency, (b) real and
imaginary parts of the electron self energy vs. energy below the Fermi level,
and (c) the resulting dispersion vs. bare quasiparticle energy.
These figures are for the Einstein model for phonons, with $\omega_E = 80$ meV
and $\lambda_E = 0.5$. See text for a discussion of the multivalued portion.}
\end{center}
\end{figure}

\begin{figure}[tb]
\begin{center}
\vskip0.2cm
\epsfig{figure=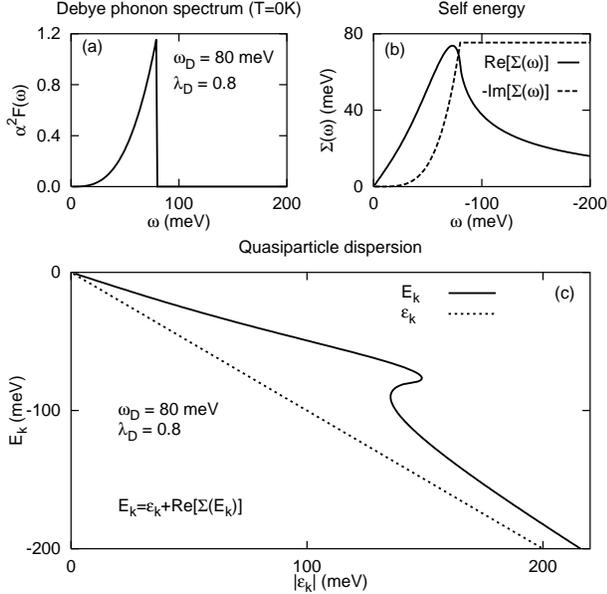,width=8cm}
\caption{Same as in Fig. 2, except for the Debye model, with
$\omega_D = 80$ meV and $\lambda_D = 0.8$.}
\end{center}
\end{figure}

In addition, in the case of the Einstein spectrum there is an unphysical singularity. 
While this is easily washed out by temperature, thermal effects still do not reconcile theory
with experiment.
Example calculations are shown in Fig. 4. {\bff We use the full 
finite temperature expression for the self energy \cite{allen82}}:
\bea
\Sigma(\omega + i\delta) = \int_0^\infty d\nu \ \alpha^2F(\nu) 
\biggl[ \,
-2\pi i \left(n(\nu) +{1 \over 2} \right) 
\nonumber \\ 
 \, \,
+\psi \left( {1 \over 2} + i{\nu - \omega \over 2\pi T} \right) -
\psi \left( {1 \over 2} - i{\nu + \omega \over 2\pi T} \right)
\, \, \biggr]
\label{self4}
\eea
where $\psi (x)$ is the digamma function and $n(\nu)$ is the Bose distribution
function.
Similarly, for the Debye
model there is a well pronounced shoulder, but still no clear kink. To determine what sort of
spectrum does lead to a kink we utilize the procedure described below.

\begin{figure}[htb]
\begin{center}
\epsfig{figure=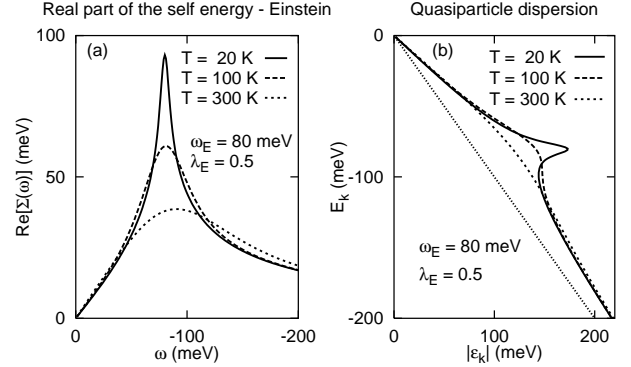,width=8cm}
\caption{Temperature dependence of (a) the real part of the self
energy, and (b) the dispersion relation for
the Einstein model shown in Fig. 2. Note that the singularity is washed out
by finite temperature.}
\end{center}
\end{figure}

As a preliminary analysis we note that the following 
function fits the data in the underdoped regime (where the kink is 
particularly prominent) reasonably well:
\begin{equation}
\Sigma_1(E_k) = \cases{
     -\lambda E_k  \phantom{aaaaaa} \mbox{if $|E_k| < \omega_D$} \cr
     -\lambda {\omega_D^2 \over E_k}  \phantom{aaaaaa} 
       \mbox{if $|E_k| > \omega_D$} \cr}.
\label{self1}
\end{equation}
{\bff This parametrization is motivated by the following considerations: we wanted an
analytical form that would contain an explicit `kink', we wanted as simple a form as
possible, and we wanted a form with which we could analytically perform the Kramers-Kronig
integration to obtain the imaginary part of the self energy.} 
{\bff We have had to use an additional fit for the high energy region, to relate
$\epsilon_k$ to $k - k_F$. For simplicity, we have used a linear fit, i.e.
$\epsilon_k = xk^\prime$, with $k^\prime = |k - k_F|$. 
Inclusion of a quadratic term (see Ref. \cite{lashell00} for a discussion
of quadratic corrections) results in a correction which is negligible. 
The proportionality constant $x$ has units of eV \AA \ and is related to 
the bare Fermi velocity ($x=\hbar v_F$).
For the underdoped sample, the proportionality constant has been determined by the
requirement that, at high energy, the full dispersion becomes the bare one (the model
self energy decreases to zero). Because the kink is washed out by increased doping, using
the same function to fit the data in the optimally doped and especially in the overdoped
regime is difficult. To overcome this problem we renormalize the momentum such that the
three experimental curves overlap in the high energy range. We then use the fit for the
underdoped data to calculate the proportionality constant $x$  for the high energy fit
for the optimally doped and overdoped sample.} 

The result of using three fitting parameters, $\lambda$ and $\omega_D$, and
the background slope $x$ is shown in Fig. 5; it is
clear that the fit is very good.
The important feature, captured by the fit, is the initial
linear dependence of energy on wavevector, followed by an abrupt change
at some characteristic frequency. {\bff In Fig. 5b we show the extracted real
and imaginary parts of the self energy from the data. We show the model fit,
the actual data once the high energy part is extracted, and the smoothened curve
used to carry out the Kramers-Kronig analysis.}
We find parameter values of 
$\lambda = 0.89$, $\omega_D = 72$ meV, and $v_F = 4.1$ eV-\AA$ /\hbar$ (= 6.2X10$^{7}$ cm/s).
Varying these parameters `by hand' results in a less
than 10 \% change in $\lambda$, for example, before the fit becomes
visually poor; we thus regard this as a rough measure of the uncertainty.

\begin{figure}[t]
\begin{center}
\epsfig{figure=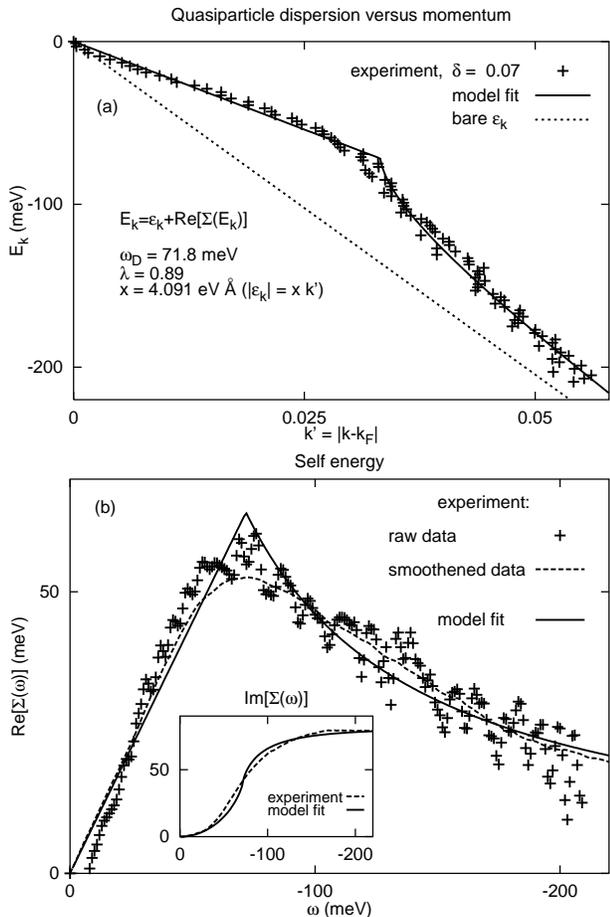,width=8cm}
\caption{(a)The fit (solid curve) to the measured dispersion data
(shown with symbols) for the underdoped case. We have used Eq. (3) with $\omega_D = 72$
meV and $\lambda = 0.89$. The dotted line shows the linear fit that results using
$x = 4.1$ eV \AA. (b) Real part of the self energy vs. energy, as extracted from the
data (symbols), the model fit (solid curve), and the smoothened fit to the data (dashed curve).
The insert shows the imaginary part obtained through Kramers-Kronig analysis.}
\end{center}
\end{figure}

The imaginary part of the self-energy can be determined from the real
part through a Kramers-Kronig integral:
\begin{equation}
\Sigma_2(\omega + i\delta) - \Sigma_2(\infty + i\delta) = {1 \over \pi}
\int_{-\infty}^{\infty} d\omega^\prime {\Sigma_1(\omega^\prime + i \delta)
\over \omega - \omega^\prime}.
\label{kk}
\end{equation}

\noindent Substituting Eq. (\ref{self1}) results in
\begin{equation}
\Sigma_2(\omega + i\delta) - \Sigma_2(\infty + i\delta) =
{2 \over \pi}\lambda \ \omega_D f\left(\omega \over \omega_D\right),
\label{imag}
\end{equation}

\noindent where $f(x) = {1 \over 2} + {(x^2 - 1) \over 4x} \ln |{1 - x \over 1 + x}|$.
With the standard approximations \cite{allen82,marsiglio02}, one can
relate the imaginary part of the self energy to the underlying
electron phonon spectrum. The required result is:
\begin{equation}
\alpha^2F(\Omega) = - {1 \over \pi} {d \over d\Omega} 
\Sigma_2(\Omega + i\delta).
\label{a2f}
\end{equation}

The result from the fit shown in Fig. 5 is plotted in Fig. 6. Note that
$f(x)$ is the same function which appears in the Hartree-Fock
calculation for the free electron gas \cite{ashcroft76}, and as is
well known, its derivative has a logarthmic singularity at the
characteristic frequency, as shown. Aside from the logarithmic singularity,
the spectrum is peculiar (as an electron phonon spectrum) in that it is 
linear at low frequencies, and has a long tail at high frequency. A model-independent
calculation (i.e. without the fit given in Eq. \ref{self1}) for each of the doping 
concentrations provided in Fig. 1 is also shown in Fig. 6. There is clearly a rounding
of the singularity \cite{note2}, although
the overall strength of the interaction, as indicated by the area under the spectrum, is
of the same magnitude. 

\begin{figure}[htb]
\begin{center}
\epsfig{figure=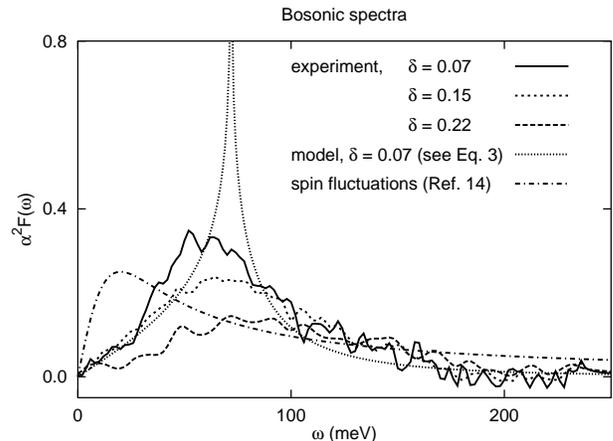,width=8cm}
\caption{The bosonic spectral functions which result from an
inversion of the data for underdoped, optimally doped, and overdoped samples
(solid, short-dashed, and long-dashed curves, respectively). Parameters for the
fits are given in Table. 1. The model based on Eq. 3, designed specifically
for the underdoped case, $\delta = 0.07$, is also shown (dot-dashed curve),
along with that expected for spin fluctuations \protect\cite{schachinger01}. In
the spin fluctuation model of \protect\cite{schachinger01} there is
considerable spectral weight at high frequencies, including weight
up to 400 meV (not shown). Note that in the inverted experimental results there
have slight negative portions, which are set to zero in the subsequent analysis
in the text.}
\end{center}
\end{figure}

Several questions arise from the results of Fig. 6. First, what sort of coupling strength
do these spectra represent, as measured by a superconducting critical temperature, assuming
that these are utilized in an Eliashberg-type analysis ? To answer this question, one would
ideally like to perform a calculation with an order parameter with d-wave symmetry.
However, for this to be meaningful one would require a series of results as shown in Fig. 1
for various directions in the Brillouin zone. Then the same inversion scheme would result
in a momentum dependent spectrum, which would then be used in a d-wave Eliashberg equation.
Since this information is lacking, we simply use an s-wave calculation, with the direct
Coulomb interaction set to zero. This would apply if the repulsion was primarily
short-range; then the d-wave symmetry would be unaffected by it. On the other hand, there is
no guarantee that the momentum dependence of the data acquired in this way would necessarily
lead to d-wave superconductivity. We proceed in this simplistic way nonetheless, and 
obtain the results for the three doping concentrations summarized in Table 1.
{\bff As judged both by the value of $T_c$ and the parameter $\lambda$, the coupling
strength decreases as the doping increases.}
Probably data from more dopant concentrations is required before one can assess how
significant this trend is, {\bff and what additional physics may causing $T_c$ to decrease
in the underdoped regime.}
{\bff We should emphasize that the main part of the analysis is done in the normal state.
Since momentum dependent data is lacking, we have tacitly assumed that the coupling strength
is independent of momentum; an important refinement will be to eventually try to extract
this momentum dependence and see how it correlates with d-wave pairing.} 

\begin{table}[htb]
\begin{center}
\vskip0.2cm
\begin{tabular}{|c|c|c|c|} 
dopant level & $\delta = 0.07$ & $\delta = 0.15$    &   $\delta = 0.22$ \\ \hline
$\lambda$      &   0.94               &      0.82                   &    0.51 \\ \hline
$\omega_{\rm ln} (meV)$  &   42       &      40                     &    38 \\ \hline
T$_c$ (K)      &   58                 &      46                     &    19 
\end{tabular}
\vskip0.2cm
\caption{
Spectral function parameters as a function of dopant concentration.
The parameters $\lambda$ and $\omega_{\rm ln}$ are determined by numerical
integrals \protect\cite{allen82} for the spectral functions shown in
Fig. 6. All indicators show a somewhat enhanced coupling at optimal doping.
} 
\end{center}
\end{table}

Another interesting question is whether these spectral functions represent phonons
or not. As pointed out in Ref. \cite{lanzara01}, the most compelling evidence 
favouring phonons is that the frequency domain
is consistent with that observed in neutron scattering. Fig. 6 does show high frequency
spectral weight, however, and one can ask whether these high frequency tails (clearly beyond the
phonon energies in these materials) rule out phonons as a possibility. To address this question
we cut off the spectrum at 100 meV, and then readjusted other parameters in the fit 
(by hand) to recover an improved fit to the data originally presented in Fig. 1. 
While the fit is never as good as the original one, we find that it is sufficiently 
good to be a plausible possibility.  Thus, unfortunately, we are unable to say 
anything very definite on this issue. It is true that spin fluctuations 
(one of the competing alternatives to phonons) are expected to have
significant spectral weight at higher frequency. An example is shown in Fig. 6,
taken from Ref. \cite{schachinger01}, which has  
considerable spectral weight extending up to 400 meV.
It is clear that significant spectral
weight exists at frequencies much higher than indicated by the experimental data; {\bff
furthermore, this particular spectrum appears to be considerably softer below the 50 meV
region than the data indicates.}
Once again more definite statements could be possible once a detailed momentum dependence of the
spectral functions is available.

{\bff Finally, we have focussed on the real part of the dispersion. One might well ask why we
didn't examine the imaginary part directly. The partial answer to this question is that there are
other (i.e. non-pairing) scattering processes which affect the imaginary part of the
self energy and not the real part (e.g. impurities --- see below).
Even more critically, as was attempted in the
original angle-resolved photoemission spectroscopy (ARPES) inversion 
work \cite{arnold91}, one might try to invert the entire 
spectral function. The difficulty is exemplified in Fig. 7, where we show an energy
distribution curve taken at $k_F$ for the underdoped sample. 
We have also plotted the energy distribution calculated with our model
fit for the self energy, Eqs. (\ref{self1}) and (\ref{imag}), additionally
including an energy resolution function \cite{fehrenbacher96} and elastic
scattering from impurities (which affects the imaginary part of the self
energy, but {\bf not} the real part).    
The fit is excellent at low energies, but there is
a clear and very large discrepancy at high energies. The origin of this discrepancy
is simply not understood at present. It may or may not represent new physics (there
may be a lack of understanding in the analysis of ARPES), but it clearly will not be
understood with the approach adopted in this paper, which focuses on the 
quasiparticle peak as an essential
feature. Mainly for this reason we felt that
an examination of the real part of the self energy (i.e. the dispersion) was the best
procedure for extracting a potential pairing interaction.}

\begin{figure}[hbt]
\begin{center}
\epsfig{figure=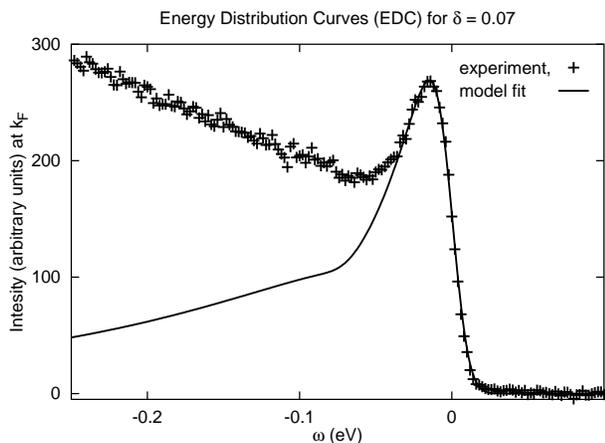,width=8cm}
\caption{The energy distribution function (spectral function times the fermi function
convoluted with an energy resolution function \protect\cite{fehrenbacher96} vs.
energy, for the underdoped sample, at $k_F$. The impurity scattering rate used was
$1/\tau = 80$ meV, and the energy broadening was 8 meV. The fit is excellent at
low energy, but clearly fails at high energy.
}
\end{center}
\end{figure}

In summary we have used the electron dispersion, as measured by ARPES, to extract an electronic
self energy (real part). Through Kramers-Kronig we are able to extract the imaginary
part of the self energy, from which an inelastic scattering spectral function is extracted in
a straightforward manner. The result is summarized in Fig. 6 for a variety of doping concentrations
in LSCO. The result is clearly compatible with phonons; the extracted coupling strength would
then be able to account for the superconductivity.
However, based on the available data we are unable to rule out, for example,
spin fluctuations as a possibility. Measurements over more doping concentrations and in
different directions in the Brillouin zone would aid considerably in narrowing down the
possibilities.

\bigskip
We would like to acknowledge Z.-X. Shen for initially encouraging us
to perform this study. We thank him, A. Lanzara and Xingjiang Zhou for the communication
of their data in electronic form, and for further discussions. We also thank an anonymous
referee for comments and questions which have improved this paper.
This work was supported by the Natural Sciences 
and Engineering Research Council (NSERC) of Canada and the Canadian
Institute for Advanced Research.



\begin{references}
\bibitem{note2} This is due to two reasons: first, the data does not, after all, 
display a perfect kink, and second, we have had to employ some smoothing to the
data to perform the Kramers-Kronig and subsequent analysis.
\end{references}
\end{document}